\documentstyle[12pt]{article}
\makeatletter

\newcommand{\PRD}[1]{{\it Phys.\ Rev.\ }{\bf D#1}}
\newcommand{\PLB}[1]{{\it Phys.\ Lett.\ }{\bf B#1}}
\newcommand{\PRC}[1]{{\it Phys.\ Rep.\ }{\bf C#1}}
\newcommand{\NPB}[1]{{\it Nucl.\ Phys.\ }{\bf B#1}}
\newcommand{\ZPC}[1]{{\it Z.\ Phys.\ }{\bf C#1}}
\newcommand{\CPC}[1]{{\it Comput.\ Phys.\ Commun.\ }{\bf #1}}
\newcommand{\hep}[1]{[{\rm hep-ph/#1}]}

\newcommand{\mrm}[1]{\mathrm{#1}}

\renewcommand{\d}{{\mrm{d}}}
\newcommand{\e}{{\mrm{e}}}

\newcommand{\p}{{\mrm{p}}}

\newcommand{\me}{m_{\e}}

\newcommand{\gtrsim}{\raisebox{-0.8mm}%
{\hspace{1mm}$\stackrel{>}{\sim}$\hspace{1mm}}}
\newcommand{\lessim}{\raisebox{-0.8mm}%
{\hspace{1mm}$\stackrel{<}{\sim}$\hspace{1mm}}}

{\end{list}}
\newcounter{enumct}

\newlength{\abstwidth}
\begin{document}
 
\sloppy

\renewcommand{\arraystretch}{1.5}

\pagestyle{empty}

\begin{flushright}
CERN--TH/96--297
\end{flushright}
 
\vspace{\fill}
 
\begin{center}
{\Large\bf 
Improving the equivalent-photon\\[1ex] 
approximation in electron--positron collisions}\\[10mm]
{\Large Gerhard A. Schuler$^a$} \\[3mm]
{\it Theory Division, CERN,} \\[1mm]
{\it CH-1211 Geneva 23, Switzerland}\\[1mm]
{ E-mail: Gerhard.Schuler@cern.ch}
\end{center}
 
\vspace{\fill}
 
\begin{center}
{\bf Abstract}\\[2ex]
\begin{minipage}{\abstwidth}
The validity of the equivalent-photon approximation for two-photon
processes in electron--positron collisions is critically examined. 
Commonly used forms to describe hadronic two-photon production are 
shown to lead to sizeable errors. An improved two-photon luminosity 
function is presented, which includes beyond-leading-logarithmic effects 
and scalar-photon contributions. Comparisons of various approximate
expressions with the exact calculation in the case of the total hadronic
cross section are given. Furthermore, effects of 
the poorly known low-$Q^2$ behaviour of the 
virtual hadronic cross sections are discussed.
\end{minipage}
\end{center}

\vspace{\fill}
\noindent
\rule{60mm}{0.4mm}

\vspace{1mm} \noindent
${}^a$ Heisenberg Fellow.

\vspace{10mm}\noindent
CERN--TH/96--297 \\
October 1996

\clearpage
\pagestyle{plain}
\setcounter{page}{1}

With the advent of LEP2, measurements of two-photon processes in a new 
domain of $\gamma\gamma$ cm energies $W$ will soon become feasible 
\cite{LEP2}.  Together with HERA measurements \cite{HERA}
of $\gamma \p$ collisions, detailed insights into the photon structure 
at the highest energies are ahead of us. 
At HERA, measurements of the scattered lepton in the luminosity 
system restricts the photon virtuality $Q^2$ in tagged $\gamma \p$ collisions
to values below about $0.02\,$GeV$^2$. This number is well below the 
typical hadronic scale $Q_0 \sim m_\rho$ and, hence, the extrapolation  
to the real photon--proton cross section is under good control. 
In contrast, two-photon processes at LEP cannot be measured at such low 
virtualities $Q_i^2$ of the two photons. Either antitagging conditions 
are imposed or the scattered leptons are measured at rather large angles 
(single- or double-tag events). In the first case, photon virtualities 
up to several GeV$^2$ are included in the data sample, while $Q^2$ 
values well above $1\,$GeV$^2$ are selected in the second case. Even the
advent of the so-called very small angle tagger will select events with 
an average $Q^2$ of still about $0.5\,$GeV$^2$ \cite{LEP2} (see also 
table~1).

In any case, the extrapolation of hadronic two-photon processes to 
zero $Q^2$ is highly non-trivial, in particular in view of the recent 
very low-$Q^2$ data from HERA \cite{lowQ}, which show a significant 
change in the $W$-dependence of the virtual $\gamma$p cross section 
as $Q^2$ exceeds $m_\rho^2$. We hasten to add that measurements 
of hadronic two-photon cross sections at non-zero photon virtualities 
might give a glimpse on the elusive QCD Pomeron \cite{LEP2,gagastar}. 
In this letter we estimate the uncertainties associated with the 
extraction of the theoretically interesting hadronic two-photon 
cross sections from the measured $\e^+\e^-$ ones and propose an 
improved equivalent-photon approximation (EPA). Let us recall that
the EPA is implemented, in one way or another, in practically all 
programmes 
\nocite{herwig,pythia,phojet,minijet,ggps1,gghv01}
[5--10]
to generate hadronic two-photon interactions.

The concept of the two-photon luminosity function ${\cal L}(\tau)$ 
arises when one relates the cross section for the scattering 
of two {\em real} photons $\sigma_{\gamma\gamma}(W^2)$ to the 
measured $\e^+\e^-$ cross section $\sigma(s)$
\begin{equation}
  \sigma(s) = \int_{\tau_{{\mrm th}}}^{1}\, 
     \d \tau\, \sigma_{\gamma\gamma}(W^2=\tau s)\, 
    {\cal L}(\tau)
\ .
\label{relation}
\end{equation}
Here $\sqrt{s}=2 E$ denotes the $\e^+\e^-$ c.m.\ energy
and $W$ the $\gamma\gamma$ c.m.\ (``hadronic'') energy. 
Usually \cite{Budnev}, ${\cal L}(\tau)$ is calculated as the product 
(or, more precisely, the convolution) of two equivalent-photon 
approximations (EPA): 
\begin{eqnarray}
  {\cal L}_{{\mrm{EPA}}}(\tau) & = & 
    \frac{1}{\tau}\,
  \int N(x_1)\, N(x_2 = \tau/x_1)\, 
    \frac{\d x_1}{x_1}
\nonumber\\
  N(x_i) & = & \frac{\alpha}{2 \pi}\, \left\{ 
  \left[ 1 + (1-x_i)^2 \right]\, \ln \frac{Q^2_{i\max}}{Q_{i\min}} \,
  - 2 m_{\e}^2 x_i^2\, \left[ \frac{1}{Q_{i\min}^2}
               - \frac{1}{Q_{i\max}^2} \right] \right\}
\nonumber\\
  Q^2_{i\min} & = & Q_{i0}^2 \equiv \frac{\me^2\ x_i^2}{1-x_i}
  \qquad\quad\,\,\, (\theta_{i\min} = 0)
\nonumber\\
              & = & (1-x_i)\ E^2\, \theta_{i\min}^2 
  \qquad (\theta_{i\min} \neq 0)
\nonumber\\
  Q^2_{i\max} & = & \min\left\{ m_{\rho}^2, (1-x_i)\ E^2\, 
         \theta_{i\max}^2 \right\}
\nonumber\\
  \max\{ x_{1\min} , \tau/x_{2\max} \} & \leq & x_1 \leq
  \min\{ x_{1\max} , \tau/x_{2\min}, 1 \}
\ .
\label{firstEPA}
\end{eqnarray}
The limits on the photon virtualities $Q_i^2$ and the scaled photon energies 
$x_i = \omega_i / E$ are determined by the experimental 
(anti-)tagging cuts on the angles $\theta_i$ and energies $E_i$ 
of the scattered electrons
$x_{i\min}  =  1 - E_{i}^{\max} / E$, 
$x_{i\max}  =  1 - E_{i}^{\min} / E$. 

It is important to realize the 
three approximations that lead to (\ref{firstEPA}) \cite{Budnev}. 
First, the $Q_i^2$ dependence of $\sigma_{TT}$, the cross section for
transverse photons, is neglected. Rather, for a hadronic cross 
section the $Q_i^2$ integrations are cut off at $Q_i^2 = m_{\rho}^2$ 
in (\ref{firstEPA}). For $Q_i^2 \lessim m_{\rho}^2$ 
the uncertainty is of the order of $Q_i^2/m_{\rho}^2$. 
Clearly, the approximation will break down for tagged events 
where $Q^2 \gtrsim m_\rho^2$.

Second, scalar-photon contributions are neglected. 
Again, for $Q_i^2 \lessim m_{\rho}^2$ 
the uncertainty is bound by $Q_i^2/m_{\rho}^2$, but 
scalar-photon contributions can potentially be large in tagged events. 

Third, the kinematics is not treated exactly. Rather it is based on the 
approximation $Q_i^2 \ll W^2$. 
For $Q_i^2 \lessim W^2$, the uncertainty is of order $Q_i^2/W^2$ and, hence,  
presumably small for the measurement of cross sections at large $W$  
($W > 5\,$GeV, say,) since the dynamics will strongly suppress $Q^2$ values
larger than a few GeV$^2$.

We now investigate the importance of the various contributions 
and propose an improved luminosity function.
Consider the reaction $\e^+(p_1) + \e^-(p_2) \rightarrow 
\e^+(k_1) + \e^-(k_2) +  X$ proceeding through the two-photon process 
$\gamma(q_1) + \gamma(q_2) \rightarrow X$, $q_i = p_i - k_i$, 
$Q_i^2 = - q_i^2$, $W^2 = (q_1 + q_2)^2$.
In general, any two-photon process is described by five non-trivial 
structure functions (two more for polarized initial electrons). 
Three of these can be expressed through the cross sections 
$\sigma_{ab}$ for scalar ($a,b=S$) and transverse photons ($a,b=T$) 
($\sigma_{ST} = \sigma_{TS}(q_1 \leftrightarrow q_2)$). 
The other two structure functions $\tau_{TT}$ and $\tau_{TS}$
correspond to transitions with spin-flip
for each of the photons (with total helicity conservation, of course). 
We emphasize that the hadronic physics is fully encoded in these structure 
functions while the connection with the measured $\e^+\e^-$ cross section
is a pure matter of QED. This connection is most transparent \footnote{%
See, for example, \cite{Budnev} where also explicit expressions for the
density matrices $\rho_i^{ab}$ of virtual photons can be found.}
if we introduce $\tilde{\phi}$, the angle between the scattering 
planes of the colliding $\e^+$ and $\e^-$ in the {\em photon} c.m.s.:
\begin{eqnarray}
 \d\sigma & = & \frac{\alpha^2}{16\, \pi^4\, Q_1^2\, Q_2^2}\, 
  \sqrt{ \frac{ (W^2+Q_1^2+Q_2^2)^2 - 4\, Q_1^2 Q_2^2 }
              { s\, (s - 4\, \me^2) } }\, \ 
         \frac{\d^3 k_1}{E_1}\,  \frac{\d^3 k_2}{E_2}
  \,  \left\{ 4\, \rho_1^{++} \rho_2^{++}\, \sigma_{TT} 
\right.
\nonumber\\ & &~ 
             + 2\, \rho_1^{++} \rho_2^{00}\, \sigma_{TS}
             + 2\, \rho_1^{00} \rho_2^{++}\, \sigma_{ST}
             + \rho_1^{00} \rho_2^{00}\, \sigma_{SS}
\nonumber\\ 
 & &~ \left. 
     + 2\, |\rho_1^{+-} \rho_2^{+-}|\, \tau_{TT}\, \cos 2\tilde{\phi}
     - 8\, |\rho_1^{+0} \rho_2^{+0}|\, \tau_{TS}\, \cos   \tilde{\phi}
  \right\}
\ .
\label{general}
\end{eqnarray}

In exact treatments of the phase space, the latter is most often 
\nocite{Budnev,Field,twogam,twogen} 
[11--14]
expressed in terms of the
virtualities $Q_i$ (or the polar angles $\theta_i$) 
and energies $\omega_i = q_i \cdot (p_1+p_2)/\sqrt{s}$  
of the photons, and of the angle $\phi$ between the planes of 
the two scattered electrons defined in the {\em laboratory} c.m.s.:
\begin{equation}
         \frac{\d^3 k_1}{E_1}\,  \frac{\d^3 k_2}{E_2}
  = \frac{2\, \pi}{s - 4\, \me^2}\, 
    \d Q_1^2\,\d Q_2^2\, \d \omega_1\, \d \omega_2 \, \d \phi
\ .
\label{phasespace}
\end{equation}
%
Note that, in general, $\tilde{\phi} \neq \phi$ and the hadronic 
energy $W$ depends non-trivially on the integration variables
\begin{equation}
  W^2 = W_A^2 + \sqrt{ \frac{W_B^4}{4\,(E^2 - \me^2)^2} }\, \cos\phi
\ ,
\end{equation}
where
\begin{eqnarray}
W_A^2 & = & 4\,\omega_{{1}}\omega_{{2}}
 - {\frac {{Q_{{2}}}^{2}\left ({E}-\omega_{{1}}\right )}
{{E}}}
 - {\frac {{Q_{{1}}}^{2}\left ({E}-\omega_{{2}}\right )}
{{E}}}
 + {\frac {{Q_{{2}}}^{2}{Q_{{1}}}^{2}}{2\,{{E}}^{2}}}
\nonumber\\
& & +{\frac {2\,{\me}^{2}\omega_{{1}}\omega_{{2}}}{{{E}}^{2}
-{\me}^{2}}}
 + {\frac {{\me}^{2}\left (2\,{E}\,{Q_{{2}}}^{2}\omega_{{1}}+
2\,{E}\,{Q_{{1}}}^{2}\omega_{{2}}+{Q_{{2}}}^{2}{Q_{{1}}}^{2}
\right )}{2\,{{E}}^{2}\left ({{E}}^{2}-{\me}^{2}\right )}}
\\
W_B^4 & = & -{Q_{{2}}}^{2}{Q_{{1}}}^{2}\left (4\,{{E}}^{2}-4\,
{E}\,\omega_{{2}}-{Q_{{2}}}^{2}\right )\left (-4\,{{E}}^{2}+4
\,{E}\,\omega_{{1}}+{Q_{{1}}}^{2}\right )
\nonumber\\
& & -4\,{Q_{{2}}}^{2}{\me}^{2}{\omega_{{1}}}^{2}\left (4\,{{E}
}^{2}-4\,{E}\,\omega_{{2}}-{Q_{{2}}}^{2}\right )
 - 4\,{Q_{{1}}}^{2}{\omega_{{2}}}^{2}{\me}^{2}\left ( 4\,
{{E}}^{2}-4\,{E}\,\omega_{{1}}-{Q_{{1}}}^{2}\right )
\nonumber\\
& & + 4\,{Q_{{2}}}^{2}{Q_{{1}}}^{2}{\me}^{2}\left ({Q_{{1}}}^{2}
+{Q_{{2}}}^{2}-8\,{{E}}^{2}+4\,{E}\,\omega_{{2}}+4\,
{E}\,\omega_{{1}}\right )
\nonumber\\
& & + 16\,{\omega_{{2}}}^{2}{\omega_{{1}}}^{2}{\me}^{4}
 + 16\,{\me}^{4}\left ({Q_{{1}}}^{2}{\omega_{{2}}}^{2}
+{Q_{{2}}}^{2}{Q_{{1}}}^{2}+{Q_{{2}}}^{2}{\omega_{{1}}}^{2}\right )
\ .
\end{eqnarray}
It is only for small $Q_1$ (and/or small $Q_2$) and small $\me$ that 
$\tilde{\phi}$ coincides with $\phi$ and $W$ becomes independent of $\phi$ 
and, hence, the azimuthal integration becomes trivial. 

In order to obtain the most general $\e^+\e^-$ cross section at 
{\em fixed} $W$, equivalently at fixed $\tau$, one needs to transform 
from one of the integration variables of (\ref{phasespace}) to $W$
\nocite{Bonneau,vermaseren,diag36,Diberder}
[15--18], for example
from $\phi$ to $W$ \cite{Bonneau}
\begin{equation}
         \frac{\d^3 k_1}{E_1}\,  \frac{\d^3 k_2}{E_2}
  = \frac{4\, \pi\, \sqrt{s}}{ \sqrt{s - 4\, \me^2}}\, 
    \d Q_1^2\,\d Q_2^2\, \d W^2\, 
   \frac{ \d \omega_1\, \d \omega_2}
  {\sqrt{-16\, \Delta_4}}
\ ,
\end{equation}
where the Gram determinant $\Delta_4$
is a quadratic function in each $\omega_i$.
Since the virtual-photon cross sections $\sigma_{ab}$ do not 
depend on $x_i$, the $\e^+\e^-$ cross section at fixed $\tau = W^2/s$ 
can be written as
\begin{eqnarray}
  \frac{\d\sigma(s,\tau)}{\d \tau} & \equiv & 
   {\cal L}(\tau,s,\me^2,{\mrm{cuts}})\, 
  \sigma_{\gamma\gamma}(W^2=\tau s)
\nonumber\\
 & = & 
  \sum_{a,b=T,S}\, 
 \int \d Q_1^2\, \d Q_2^2 \, J_{ab}(\tau,Q_1^2,Q_2^2;s,\me^2;{\mrm{cuts}})\, 
  \sigma_{ab}(W^2=\tau s,Q_1^2,Q_2^2)
\ ,
\label{mini}
\end{eqnarray}
where
\begin{equation}
  J_{ab} = \frac{1}{\pi}\,
   \frac{ \d \omega_1\, \d \omega_2}{\sqrt{ -16\, \Delta_4 } }\,
   \frac{\alpha^2}{2\, \pi^2\, Q_1^2\, Q_2^2}\, 
   \frac{s \sqrt{X}}{ s - 4\, m^2 }\, 
  \rho_1^{ab}\,  \rho_2^{ab}
\ .
\end{equation}
Here we have neglected the $\tilde{\phi}$-dependent terms, which were
significant only when experimental situations with asymmetric cuts 
on the two scattered leptons would not be averaged.
If one is interested in only the total cross section at fixed $W$, 
analytic results can be obtained for the functions $J_{ab}$. Indeed, 
an explicit expression for $J_{TT}$ has been derived in \cite{Bonneau}. 
However, in experimental measurements one always applies cuts 
on the angles and energies of scattered electrons, in which case 
exact results for $J_{ab}$ cannot be obtained analytically. 
On the other hand, accurate numerical results (better than $1\%$, say) 
are not easily obtained since (\ref{mini}) involves four non-trivial 
integrations.

For applications at LEP, where the experimental interest is
focused on two-photon physics at high $W$, one can proceed further 
analytically by expanding in $Q_i^2/W^2$. We start from the observation
that, for $\me^2/s \ll 1$, $\me^2/W^2 \ll 1$, and when at least 
one $Q_i$ is small compared with $W$, the $\omega$ integration measure 
can be replaced by the approximate expression with a Dirac delta distribution:
\begin{equation}
 \frac{1}{\pi}\,
   \frac{ \d \omega_1\, \d \omega_2}{\sqrt{ - 16\,\Delta_4 } }
       = \frac{1}{s}\, 
 \left\{ 
   \frac{ \d \omega_1\, \d \omega_2}{ 4} 
   \delta \left( \omega_1 \omega_2 - W^2/4 \right)
  + {\cal O}(\me^2/s, \me^2/W^2, Q_i^2/W^2 ) \right\}
\ .
\label{delta}
\end{equation}
The approximation (\ref{delta}) is justified for $W \gg m_\rho$, 
since hadronic two-photon cross sections $\sigma_{ab}$ vanish quickly 
for $Q_i^2 \gtrsim m_\rho^2$. 
This yields
\begin{equation}
x_1\, \tau\, \frac{\d {\cal L}}{\d x_1} = 
  \int\, \d Q_1^2\,      \int\, \d Q_2^2\,   
  \sum_{a,b=T,S}\, f_a(x_1,Q_1^2)\, f_b(x_2,Q_2^2)\, 
      \frac{\sigma_{ab}(W^2,Q_1^2,Q_2^2)}
           {\sigma_{\gamma\gamma}(W^2)}
\ ,
\label{firstsimple}
\end{equation}
where
\begin{eqnarray}
 f_T(x,Q^2) & = & \frac{\alpha}{2\ \pi}\, \left\{ 
  \frac{ 1 + (1-x)^2}{Q^2} - \frac{2\ \me^2\ x^2}{Q^4} \right\}
  \equiv f_{{\mrm{LLA}}} +  f_{{\mrm{NL}}} 
\nonumber\\
 f_S(x,Q^2) & = & \frac{\alpha}{\pi}\, \frac{1-x}{Q^2}
\ .
\label{fTSdef}
\end{eqnarray}
The $Q_i^2$-integration limits are determined by the experimental 
(anti-)tagging cuts on the angles of the scattered electrons 
$\theta_{i\min,\max}$. For $\me \ll \omega_i$ we find
\begin{eqnarray}
  Q^2_{i\min} & = & Q_{i0}^2 + 4 E^2 (1-x_i) \sin^2\frac{1}{2}\theta_{i\min}
  \approx Q_{i0}^2 + E^2 (1-x_i) \theta_{i\min}^2
\nonumber\\
  Q^2_{i\max} & = & Q_{i0}^2 + 4 E^2 (1-x_i) \sin^2\frac{1}{2}\theta_{i\max}
  \approx Q_{i0}^2 + E^2 (1-x_i) \theta_{i\max}^2
\ .
\label{Qimin}
\end{eqnarray}

Note the inclusion of the usually neglected term 
$Q_{i0}^2 = \me^2\, x_i^2/(1-x_i)$ in the upper 
limit of $Q^2$. Its presence improves the behaviour for 
$x_i \rightarrow 1$. In the limit of small maximum scattering angle 
$\theta_{1\max}$ and $\theta_{1\min}=0$, 
the $Q_1^2$-limits (\ref{Qimin}) coincide with the results 
quoted in \cite{Ridolfi} found from an analysis of $\e\p$ collisions. 

In order to further proceed analytically we make use of a second observation. 
At $Q_i^2 \ll W^2$ one can, to a very good approximation, assume 
factorization of the $W$ and $Q_i$ dependences:
\begin{equation}
  \sigma_{ab}(W^2,Q_i^2) = h_a(Q_1^2)\, h_b(Q_2^2)\, 
\sigma_{\gamma\gamma}(W^2) 
\ .
\label{sigapprox}
\end{equation}
%
Then (\ref{mini}) can be reduced to the single product of a ``true'' 
two-photon luminosity function and the real $\gamma\gamma$ cross section, 
the former given by
\begin{equation}
  {\cal L}  =   
\frac{1}{\tau}\, \int \frac{\d x_1}{x_1}\,
  \prod_{i=1}^{2} 
  \left\{ \int \d Q_i^2\, \left[ f_T(x_i,Q_i^2)\, h_T(Q_i^2)
                              +  f_S(x_i,Q_i^2)\, h_S(Q_i^2)
   \right] \right\}
\ .
\label{Lumii}
\end{equation}
Equation (\ref{Lumii}) defines the improved EPA. 
We emphasize that, in general, the luminosity function does not factorize 
into the product of two equivalent-photon approximations; the approximation 
$Q_i^2 \ll W^2$ is essential.

We now present our results for four typical experimental situations, 
antitag events (both electrons in the antitag region), low-$Q^2$ 
and high-$Q^2$ single-tag events (one electron in the antitag region and
the other in the tag region), and low-$Q^2$ double-tag events. We 
choose $\sqrt{s}=130\,$GeV, $W=10\,$GeV and define the regions
in table~1.
\begin{table}[bht]
\begin{center}
\begin{tabular}{|r|r|r|r|r|l|r|}
\hline
Tag region 
& $E_{1\min}$ &  $E_{1\max}$
& $\theta_{1\min}$ & $\theta_{1\max}$ 
& $Q_{1\min}^2$  & $Q_{1\max}^2$ 
\\ \hline
Antitag & $0$ & $65$ & $0$ & $1.43$ & $0.572\times10^{-12}$ & $5.26$
\\ \hline
High-$Q^2$ tag & $30$ & $65$ & $1.55$ & $3.67$ & $2.85$ & $34.7~$
\\ \hline
Low-$Q^2$ tag & $30$ & $65$ & $0.28$ & $0.68$ & $0.0931$ & $1.19$
\\ \hline
\end{tabular}
\caption[]{The cuts on the scattered electrons used in this paper. 
For convenience, the corresponding $Q^2$-ranges are also given
($\sqrt{s}=130\,$GeV, $W=10\,$GeV; energies in GeV, angles in degrees, 
and squared momentum transfers in GeV${}^2$).}
\end{center}
\end{table}

In order to appreciate the uncertainty associated with the poorly known
$Q^2$ fall off of the hadronic cross sections, we consider three models. 
The first follows from a parametrization \cite{Bezrukov}
of the $\gamma^* \p$ cross section calculated in a model of 
generalized vector-meson dominance (GVMD):
\begin{eqnarray}
  h_T(Q^2) & = & r\, P_1^{-2}(Q^2) + (1-r)\, P_2^{-1}(Q^2) 
\nonumber\\
  h_S(Q^2) & = & \xi\, \left\{ r\, \frac{Q^2}{m_1^2}\, P_1^{-2}(Q^2)
       + (1-r)\, \left[ \frac{m_2^2}{Q^2} \, \ln P_2(Q^2) 
       -  P_2^{-1}(Q^2) \right] \right\}
\nonumber\\
  P_i(Q^2) & = & 1 + \frac{Q^2}{m_i^2}
\ ,
\label{GVDMform}
\end{eqnarray}
where we take $\xi=1/4$, $r=3/4$, $m_1^2 = 0.54\,$GeV$^2$ 
and $m_2^2 = 1.8\,$GeV$^2$.

The second model \cite{Sakurai} adds a continuum contribution to simple 
(diagonal, three-mesons only) vector--meson dominance (VMDc):
\begin{eqnarray}
  h_T(Q^2) & = & \sum_{V=\rho,\omega,\rho}\, r_V\, 
      \left(\frac{m_V^2}{m_V^2 + Q^2}\right)^2
  + r_c\, \frac{m_0^2}{m_0^2 + Q^2}
\nonumber\\
  h_S(Q^2) & = & \sum_{V=\rho,\omega,\rho}\, \frac{\xi\, Q^2}{m_V^2}\, 
      r_V\,  \left(\frac{m_V^2}{m_V^2 + Q^2}\right)^2
\ ,
\label{VMDcform}
\end{eqnarray}
where
$r_\rho=0.65$, $r_\omega = 0.08$, 
$r_\phi = 0 .05$, and $r_c = 1 - \sum_V r_V$. 
Since photon-virtuality effects are often estimated by using a simple 
$\rho$-pole 
only, we consider also the model defined by ($\rho$-pole):
\begin{eqnarray}
  h_T(Q^2) & = & \left(\frac{m_\rho^2}{m_\rho^2 + Q^2}\right)^2
\nonumber\\
  h_S(Q^2) & = & \frac{\xi\, Q^2}{m_\rho^2}\, 
      \left(\frac{m_\rho^2}{m_\rho^2 + Q^2}\right)^2
\ .
\label{rhoform}
\end{eqnarray}

In all cases, the $Q_i^2$ integrations can be performed analytically so
that one is left with only a single one-dimensional numerical integration. 
A Fortran program is available on request from the author.

\begin{table}
\begin{center}
\begin{tabular}{|l||c|c||c|c|}
\hline
Tag region & \multicolumn{2}{|c||}{Antitag [$\times 10^{-2}$]}
     & \multicolumn{2}{|c|}{Single low-$Q^2$ tag [$\times 10^{-4}$]}
\\ \hline
Model & exact & approx.\ & exact & approx.\
\\ \hline
 GVMD & $105~$ & $105~$ & $608$ & $610$ 
\\ \hline
 VMDc & $105~$ & $105~$ & $622$ & $624$
\\ \hline
 $\rho$-pole          & $102$ & $102~$ & $541$ & $543$
\\ \hline
 $\rho$-pole, $h_S=0$ & $~99.5$ & $~99.5$ & $493$ & $495$
\\ \hline
 $\rho$-pole, $h_S=f_{{\mrm{NL}}}=0$ & & $110~$ & & $522$
\\ \hline
 EPA (\ref{firstEPA}) &  & $112~$ & & $974$
\\ \hline
 PDG (\ref{PDGform}) & & $114~$ & & ---
\\ \hline
\hline
Tag region 
 & \multicolumn{2}{|c||}{Single high-$Q^2$ tag [$\times 10^{-5}$]}
 & \multicolumn{2}{|c|}{Double low-$Q^2$ tag [$\times 10^{-5}$]}
\\ \hline
Model & exact & approx.\ & exact & approx.\
\\ \hline
 GVMD & $789~$ & $851$ & $355$ & $357$ 
\\ \hline
 VMDc & $696~$ & $749$ & $370$ & $373$
\\ \hline
 $\rho$-pole          & $267~$ & $286$ & $290$ & $292$
\\ \hline
 $\rho$-pole, $h_S=0$ & $~94.9$ & $101$ & $246$ & $247$
\\ \hline
 $\rho$-pole, $h_S=f_{{\mrm{NL}}}=0$ & & $106$ & & $247$
\\ \hline
 EPA (\ref{firstEPA}) &  & $0$ & & $862$
\\ \hline
\end{tabular}
\end{center}
\caption[]{The two-photon luminosity ${\cal L}$ at $\sqrt{s}=130\,$GeV 
and $W=10\,$GeV for various models of the two-photon cross sections 
calculated exactly by integrating (\ref{mini}), in the improved EPA 
(\ref{Lumii}), and in the two commonly used EPA (\ref{firstEPA})
and (\ref{PDGform}).}
\end{table}
Table~2 gives the results for the three models 
obtained from the improved EPA (\ref{Lumii}) 
in comparison with the exact results 
obtained\footnote{Details of the exact integration (\ref{mini})
will be described elsewhere \cite{GAS}.} from (\ref{mini}).
Table~2 contains also the results of  
four approximations, two of which are frequently used EPAs: 
(i) the $\rho$-pole model without scalar-photon contributions, i.e.\ 
$h_S(Q_i^2)=0$; 
(ii) the $\rho$-pole model in the leading-logarithmic approximation, i.e.\ 
neglecting, besides scalar-photon contributions, also the $\me^2/Q^4$ term 
in $f_T$ ($f_{{\mrm{NL}}} = 0$ in (\ref{fTSdef}));
(iii) the EPA defined in (\ref{firstEPA}) \cite{Budnev}; 
and (iv) the EPA obtained by integrating (\ref{firstEPA}) with 
logarithmic accuracy \cite{Budnev}: 
\begin{eqnarray} 
  {\cal L}_{PDG} & = & \frac{1}{\tau}\, 
      \left(\frac{\alpha}{\pi}\right)^2\, 
  \left[ \left( \ln\frac{ Q_m^2}{\tau\, \me^2} - 1 \right)^2 \, 
        f(\tau,\tau_m) - \frac{1}{3}\, \ln^3 \frac{1}{\tau_m}
  \right]
\nonumber\\
 f(x,y) & = & \left( 1 + \frac{x}{2} \right)^2 \, \ln\frac{1}{y}
  - 2\, \sqrt{\frac{x}{y}} \left( 1 + \frac{x}{2} \right)
  (1 - y) + \frac{x}{2y}\, \left( 1 - y^2 \right)
\nonumber\\
  Q_m & = & \min\left\{ m_\rho, \theta_{\max}\, E \right\}
 \qquad ; \qquad \tau_m = \max\left\{ \frac{4\, \omega_{\min}^2}{W^2}, 
         \tau \right\}
\ .
\label{PDGform}
\end{eqnarray}
For $W^2 \geq 4\, E\omega_{\min}$, (\ref{PDGform}) reduces to the expression
quoted by the PDG \cite{PDG}. 

The fastest (in terms of CPU time) estimate is, of course, obtained 
with the closed expression (\ref{PDGform}) and is accurate to about $10\%$ 
for the antitag case. Clearly, (\ref{PDGform}) cannot be used for 
single-tag cases, but also the double-tag case is beyond its validity. 
The evaluation of the improved EPA  (\ref{Lumii}) or the standard EPA 
(\ref{firstEPA}) is still very fast and, definitely, much shorter than 
the evaluation of the exact result (\ref{mini}). The result of the standard
EPA (\ref{firstEPA}) is slightly better than that of (\ref{PDGform}) 
for the antitag case. 
More importantly, (\ref{firstEPA}) can also be applied to the 
low-$Q^2$-tag cases, at least for an order-of-magnitude estimate: 
the results for the single and double low-$Q^2$-tag cases are 
overestimated by factors of $1.6$ and $2.4$, respectively. 
Results for high-$Q^2$ tags cannot be obtained 
since the lower-$Q^2$ limit exceeds $m_{\rho}^2$. 
An additional drawback of the standard EPA is the fact that one cannot
investigate uncertainties in the two-photon luminosity arising from the 
poorly known low-$Q^2$ behaviour of the virtual hadronic cross sections.

As can be seen from table~2, the improved two-photon luminosity function 
(\ref{Lumii}) works extremely well in the single-tag and antitag cases. 
The accuracy is better than $1\%$ and the differences compared to the 
exact results are smaller than uncertainties associated with the 
low-$Q^2$ extrapolation of the hadronic cross sections. Even for the 
high-$Q^2$ case, the accuracy stays better than $10\%$.

The effect of neglecting the non-logarithmic term proportional to
$\me^2/Q_i^2$ in $f_T$ is as large as about $10\%$ for the antitag case. 
The importance of this term clearly diminishes with increasing
average $Q_i^2$. In the case of double low-$Q^2$ tag it becomes
essentially negligible. 

The differences between the two models GVMD and VMDc is rather small: 
in the low-$Q^2$-tag cases less than $2\%$ and in the high-$Q^2$-tag
case about $6\%$.  This result is not surprising as the two models
are quite similar. Yet, the uncertainty in the extraction of
real $\gamma\gamma$ cross section, particularly in tagged events, 
can well be much bigger. Differences of about $7\%$ ($12\%$) are found
between the VMDc and $\rho$-pole models for the single (double) 
low-$Q^2$-tag cases. The high-$Q^2$-tag results of these two models
even differ by more than a factor of $2$! These differences can be 
traced back to the extra monopole factor (continuum term) in the 
transverse-photon part of the VMDc model. 
On top of that, there is a similarly large effect of scalar photons.

A correct understanding of the hadronic 
cross sections at low photon virtualities is hence indispensable for
a precision measurement of the total hadronic $\gamma\gamma$ cross section. 
The argument can, of course, be turned around: given the large sensitivity
to the low-$Q^2$ behaviour, and provided the total cross section can be
measured in various modes (antitag, low- and high-$Q^2$ single- and 
double-tag events) the transition region towards zero $Q^2$ can be 
investigated at LEP. It goes without saying that with only minor 
modifications, the above-proposed improved luminosity function can
also be used for other hadronic reactions, for example high-$p_T$ 
jet production.\hfill\\[2ex]
\noindent
{\it Acknowledgements}\hfill\\
I thank F.\ Berends, M.\ Seymour, and D.\ Summers for useful discussions.


\end{document}